\pdfoutput=1 % only if pdf/png/jpg images are used
\documentclass{JINST}

\usepackage{amsmath}
\usepackage{subfigure}
\usepackage{booktabs}
\usepackage{lineno}

\title{Development of a Depleted Monolithic CMOS Sensor in a 150 nm CMOS Technology for the ATLAS Inner Tracker Upgrade}

\author{T.~Wang$^a$\footnote{Corresponding author}, P.~Rymaszewski$^a$, M.~Barbero$^b$, Y.~Degerli$^c$, S.~Godiot$^b$, F.~Guilloux$^c$, T.~Hemperek$^a$, T.~Hirono$^a$, H.~Kr{\"u}ger$^a$, J.~Liu$^b$, F.~Orsini$^c$, P.~Pangaud$^b$, A.~Rozanov$^b$ and N.~Wermes$^a$\\
\llap{$^a$}Physikalisches Institut der Universit{\"a}t Bonn,\\Nussallee 12, Bonn, Germany\\
\llap{$^b$}Centre de Physique des Particules de Marseille,\\Avenue de Luminy 163, Marseille, France\\
\llap{$^c$}Institut de Recherche sur les lois Fondamentales de l'Univers,\\CEA - Saclay, France\\
E-mail: \email{t.wang@physik.uni-bonn.de}}

\abstract{The recent R~\& D focus on CMOS sensors with charge collection in a depleted zone has opened new perspectives for CMOS sensors as fast and radiation hard pixel devices. These sensors, labelled as depleted CMOS sensors (DMAPS), have already shown promising performance as feasible candidates for the ATLAS Inner Tracker (ITk) upgrade, possibly replacing the current passive sensors. A further step to exploit the potential of DMAPS is to investigate the suitability of equipping the outer layers of the ATLAS ITk upgrade with fully monolithic CMOS sensors. This paper presents the development of a depleted monolithic CMOS pixel sensor designed in the LFoundry 150~nm CMOS technology, with the focus on design details and simulation results.}


\begin{document}
%\setpagewiselinenumbers
%\linenumbers

\section{Introduction}
\label{sec:intro}

Standard CMOS pixel sensors, integrating the sensing volume and the read-out electronics on the same substrate, have proven to be high precision tracking and vertexing devices in high energy particle physic experiments, thanks to their fine granularity and low material. In addition, the use of commercial technologies makes them very suitable for covering large area in trackers due to their relative low cost. However, existing CMOS detectors in operation~\cite{HFT_Greiner_2015}, or under construction~\cite{ALPIDE_Mager_2016}, still rely in part or even substantially on diffusion for charge collection, which in turn leads to limited speed and radiation hardness, and therefore are not suitable for the extreme rate and radiation environment like ATLAS. A recent trend of exploiting the high-voltage/high-resistivity add-ons of CMOS technologies to increase the depletion of the sensing volume has enabled fast charge collection by drift in CMOS sensors~\cite{HVCMOS_for_ATLAS_Ivan_2012,ESPROS_Obermann_2015,LF_Hirono_2016}. This relatively new sensor family, labeled as \emph{depleted CMOS pixel} or \emph{depleted monolithic active pixel} (DMAPS), is currently under study, targeting high performance and cost efficient CMOS detectors for the ATLAS Inner Tracker (ITk) upgrade in the High-Luminosity LHC (HL-LHC) era. Depleted CMOS pixels, that have survived the radiation levels approaching the harshest requirements of the ITk upgrade, have already been reported~\cite{HVCMOS_Liu_2015,HVCMOS_Ristic_2015}.

So far, the reported depleted CMOS pixels are either passive sensors or active ones with only first stage(s) of front-end (FE) electronics integrated in the pixel. The data readout and processing rely on a readout chip, bump bonded or capacitively coupled via glue bonds to the sensor~\cite{CCPD_Ivan_2009}. These sensors are attractive alternatives for the sensing layers of hybrid detectors, enabling the cost reduction and introducing intelligence like sub-pixel address encoding in the diode array~\cite{HVCMOS_Liu_2015}. A more ambitious step to explore the potential of depleted CMOS pixels is to develop a fully monolithic sensor with fast standalone readout. And it may be suitable for the outer layers of the ATLAS ITk upgrade, where the performance requirements are less stringent in comparison to the inner layers while the cost consideration becomes important. 

This work presents a depleted monolithic CMOS pixel sensor design in the LFoundry 150~nm CMOS technology, namely LF-Monopix01. In Section~\ref{sec:LFRD}, a short overview of R~\& D on depleted CMOS pixels in the mentioned technology is given. Then, the design challenges of a monolithic sensor are addressed in Section~\ref{sec:ChallengeMonopix}. In Section~\ref{sec:DesignMonopix}, the design details of LF-Monopix01 are described, followed by the simulation results in Section~\ref{sec:Sim}. The conclusions are given in Section~\ref{sec:Con}.

\section{Overview of R~\& D on depleted CMOS pixels in the LFoundry 150~nm technology}
\label{sec:LFRD}

\subsection{Implementation concept}
\label{sec:LFTech}

\begin{figure}[htbp]
  \centering
  \includegraphics[width=.55\textwidth]{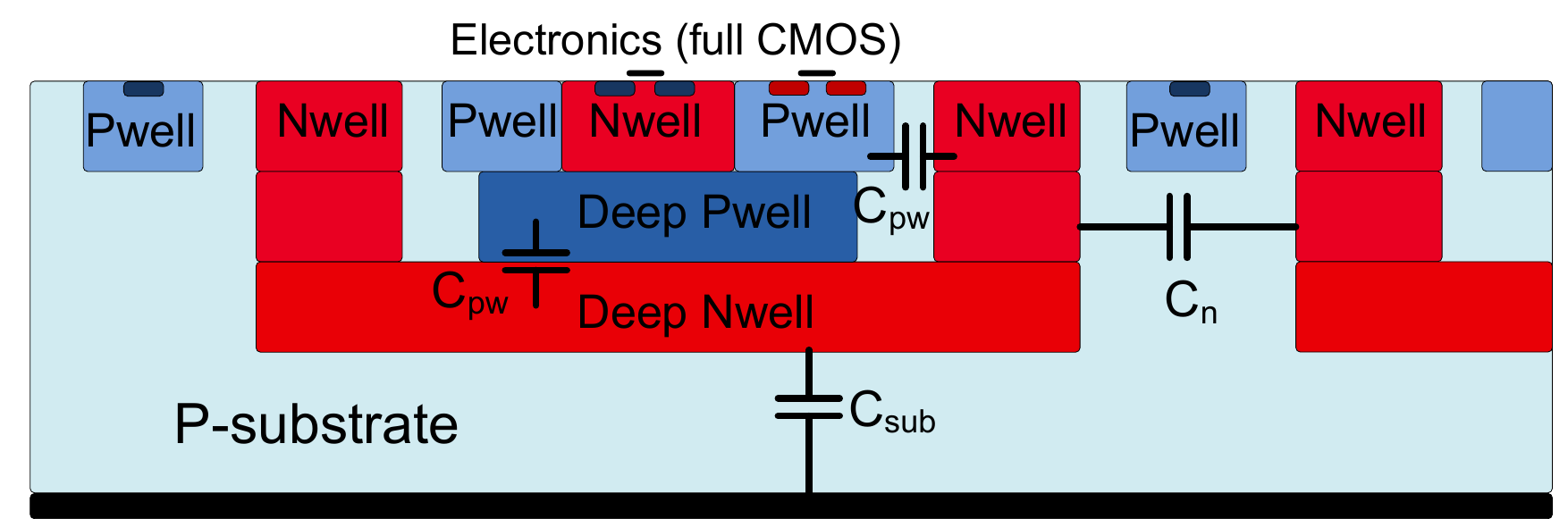}
  \caption{Schematic cross-section of a depleted CMOS pixel in LFoundry 150~nm CMOS technology. The various capacitance contributions to the sensing node are indicated by respective symbols.}
  \label{fig:Tech}
\end{figure}

A simplified schematic cross-section of a typical depleted CMOS pixel cell developed in the LFoundry 150~nm CMOS technology is shown in Figure~\ref{fig:Tech}. A high resistivity ($\sim$~2~k$\Omega \cdot$cm) P-type substrate is used for implementing the sensor. The charge collection node is formed by a deep N-well, and the depletion of the sensing volume is achieved by applying a high negative voltage to the P-substrate. The in-pixel readout electronics is implemented inside the deep N-well. With a deep P-type layer, the N-well hosting the PMOS transistors can be isolated from the deep N-well, allowing for full CMOS capability in the pixel. Wafer thinning and backside implantation are also possible after the CMOS processing, thus a slim sensor biased via backside can be achieved, yielding a strong electric field and uniform charge collection inside the sensor.

\subsection{Demonstrated capabilities}
\label{sec:LFCCPD}

The capabilities of depleted CMOS pixels implemented in the LFoundry 150~nm technology were verified by the measurement of a previous prototype~\cite{LF_Hirono_2016,LF_Rymaszewski_2016}. A depletion depth over 160~$\mu$m can be achieved with a reverse bias voltage of 110~V, confirming the high resistivity of the substrate. In-beam timing measurements show that the fraction of in-time (<~25~ns) hits is 91\% with a high threshold of 2600~e$^{-}$, and a very low threshold of 190~e$^{-}$ yields an in-time hit fraction of 79\%. The lower in-time efficiency of low threshold is due to the small amount of charge collected by the pixel resulting from charge sharing. Chip samples irradiated up to 50~Mrad of TIDs and $1 \times 10^{15}~\text{n}_{\text{eq}}\,\text{cm}^{-2}$ of neutron fluences have survived without substantial performance loss. These results indicate that depleted CMOS pixels developed in the LFoundry 150~nm technology may satisfy the timing and radiation requirements of the outer layers of the ATLAS ITk upgrade, provided that the front-end electronics is further optimized in terms of speed and digital readout. This has greatly encouraged the design of a monolithic sensor in the same technology.

\section{Design challenges of depleted monolithic CMOS pixels}
\label{sec:ChallengeMonopix}

The main consequence of implementing the sensor concept described in Section~\ref{sec:LFTech} is a non-negligible amount of increase in the detector capacitance, which in turn degrades the noise performance and the speed of the front-end electronics. The contributions of capacitance to the sensing node can be seen in Figure~\ref{fig:Tech}. Apart from the backplane capacitance C$_{\text{sub}}$ and inter-pixel capacitance C$_{\text{n}}$, which also exist in a standard n-in-p planar sensor, depleted CMOS pixels have another major contribution from the capacitance C$_{\text{pw}}$ between the collection well and the P-well hosted inside. Because the P-well sits in close vicinity to the collection well, the contribution of C$_{\text{pw}}$ is not negligible. As a matter of fact, it tends to become dominant for a monolithic sensor where complex in-pixel electronics is required. Moreover, a large C$_{\text{pw}}$ may introduce serious cross-talk, since the electronic substrate is directly coupled to the sensing node via C$_{\text{pw}}$. 

Therefore, the monolithic design calls for front-end electronics optimized for the large input capacitance. Additionally, low noise digital logic is mandatory for the in-pixel circuitry and the layout must be carefully designed to ensure good protection for the electronic substrate.

\section{Design of LF-Monopix01}
\label{sec:DesignMonopix}

The LF-Monopix01 is a fast readout monolithic CMOS sensor, and it employs the column drain architecture as used in ~\cite{ColumnRO_Mandelli_2002,FEI3_Ivan_2006}. The following section presents the design details.

\subsection{Pixel}
\label{sec:Pix}

\begin{figure}[!ht]
  \begin{center}
    \subfigure[]{
      \includegraphics[width=0.98\textwidth]{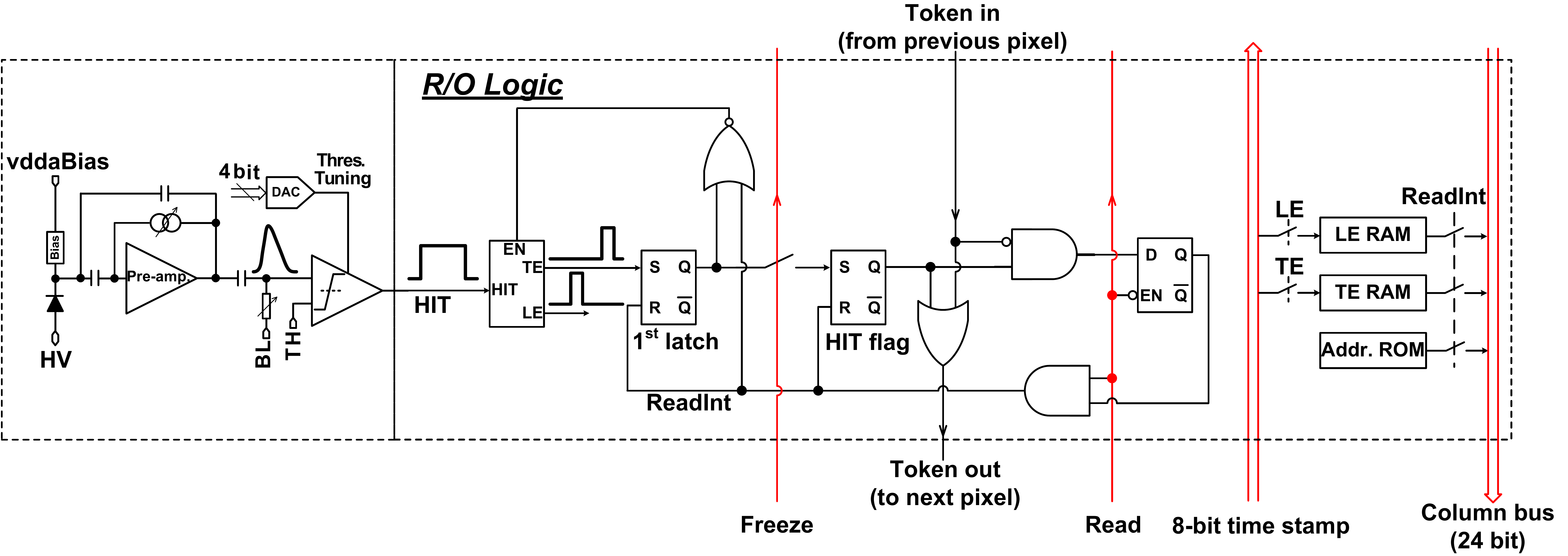}
      \label{fig:PixMono}
    }
    \subfigure[]{
      \includegraphics[width=0.45\textwidth]{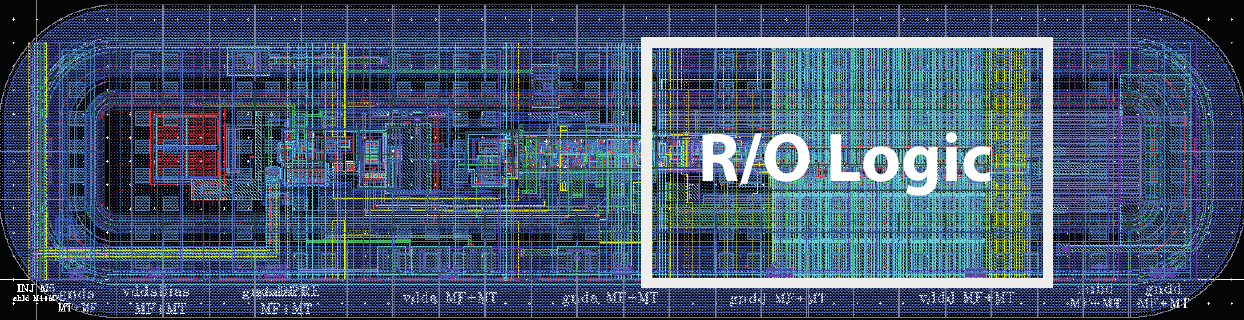}
      \label{fig:PixMonoLayout}
    }
    \subfigure[]{
      \includegraphics[width=0.27\textwidth]{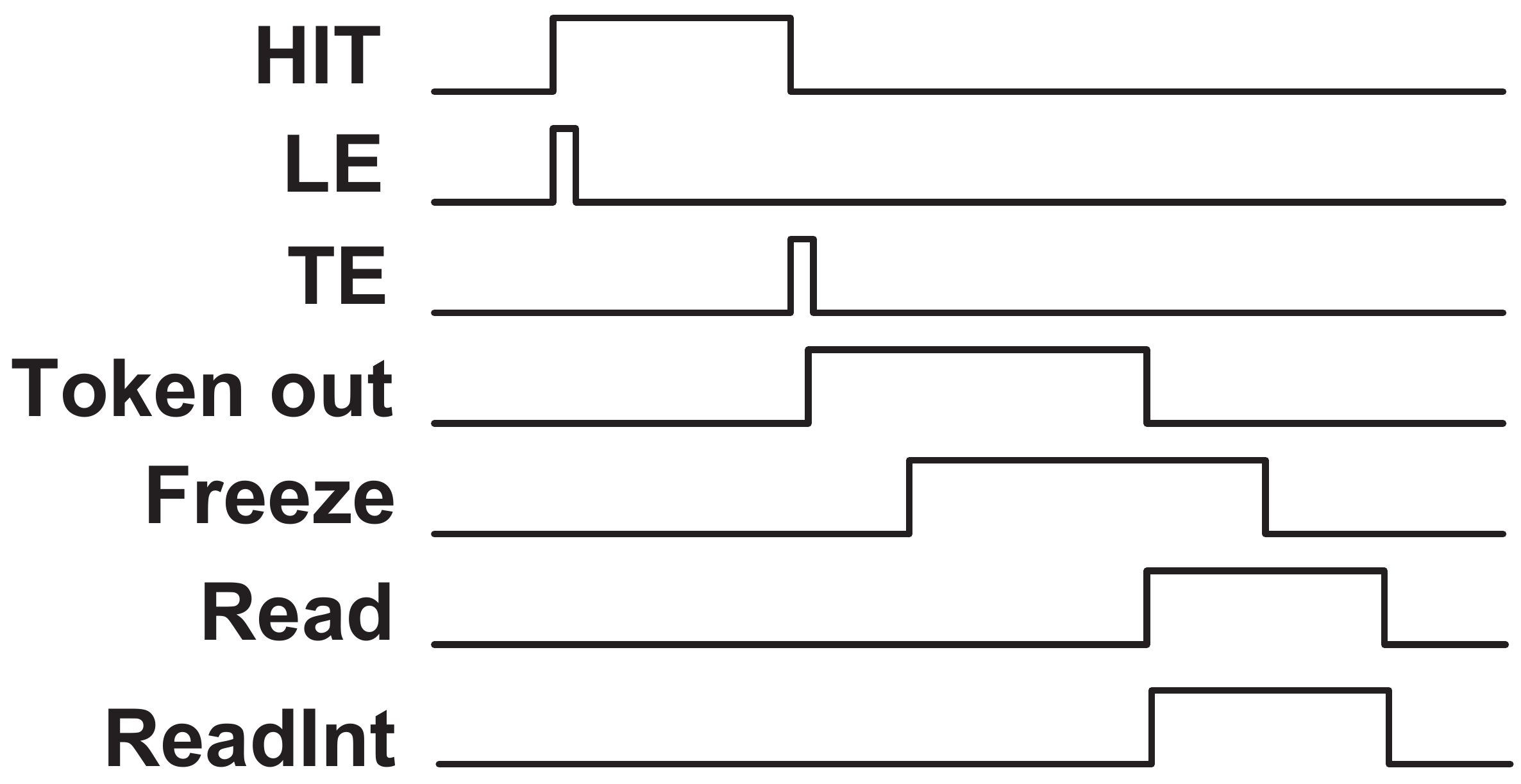}
      \label{fig:Timing}
    }
    \caption[]{\subref{fig:PixMono} Block diagram and \subref{fig:PixMonoLayout} layout of a pixel with integrated read-out logic; \subref{fig:Timing} The timing diagram for the read-out logic.}
    \label{fig:MonoPix} 
  \end{center}
\end{figure}

The pixel size of LF-Monopix01 is 50~$\times$~250~$\mu$m$^{\text{2}}$. The block diagram as well as the layout of a typical pixel in LF-Monopix01 are depicted in Figure~\ref{fig:MonoPix}. 

\paragraph{Working Principle} The signal processing starts from the pre-amplifier, which is AC coupled to the sensing diode. With the capacitive feedback, the pre-amplifier converts the charge signal to voltage signal. The output voltage signal is filtered by a baseline restoration stage, and then sent to a discriminator with the threshold adjustable by a 4-bit in-pixel DAC. The discriminator output triggers the read-out logic, whose working principle can be described with the help of the timing diagram shown in Figure~\ref{fig:Timing}. Two short pulses are generated, respectively, at the leading edge (LE) and trailing edge (TE) of the discriminator output. These two pulses strobe the 8-bit timing information into two RAM cells. The TE pulse also sets the \emph{HIT flag} register, which inserts a \emph{Token} that can propagate to the column end through a chain of \emph{OR} logic. The output \emph{Token} is received by a read-out controller (not shown in Figure~\ref{fig:PixMono}), which sends back the \emph{Freeze} signal to the column. The \emph{Freeze} signal makes sure that new hits won't disturb the read-out process, while these later hits can still be recorded by a first latch. The read-out controller then generates the \emph{Read} signal, which is selected by a priority network so that only the topmost hit pixel has its internal \emph{ReadInt} signal enabled. The \emph{ReadInt} signal allows the hit information (8-bit LE time stamp, 8-bit TE time stamp and 8-bit pixel address) from the in-pixel memories to be driven to the column data bus for readout. Since the \emph{ReadInt} signal also clears the hit flag register, the priority network ripples down to find the next hit pixel, if any, once the \emph{Read} signal goes back to zero. This read-out cycle continues by sending repetitively the \emph{Read} pulse until this frozen column is "drained out".

\begin{figure}[!ht]
  \begin{center}
    \subfigure[]{
      \includegraphics[width=0.4\textwidth]{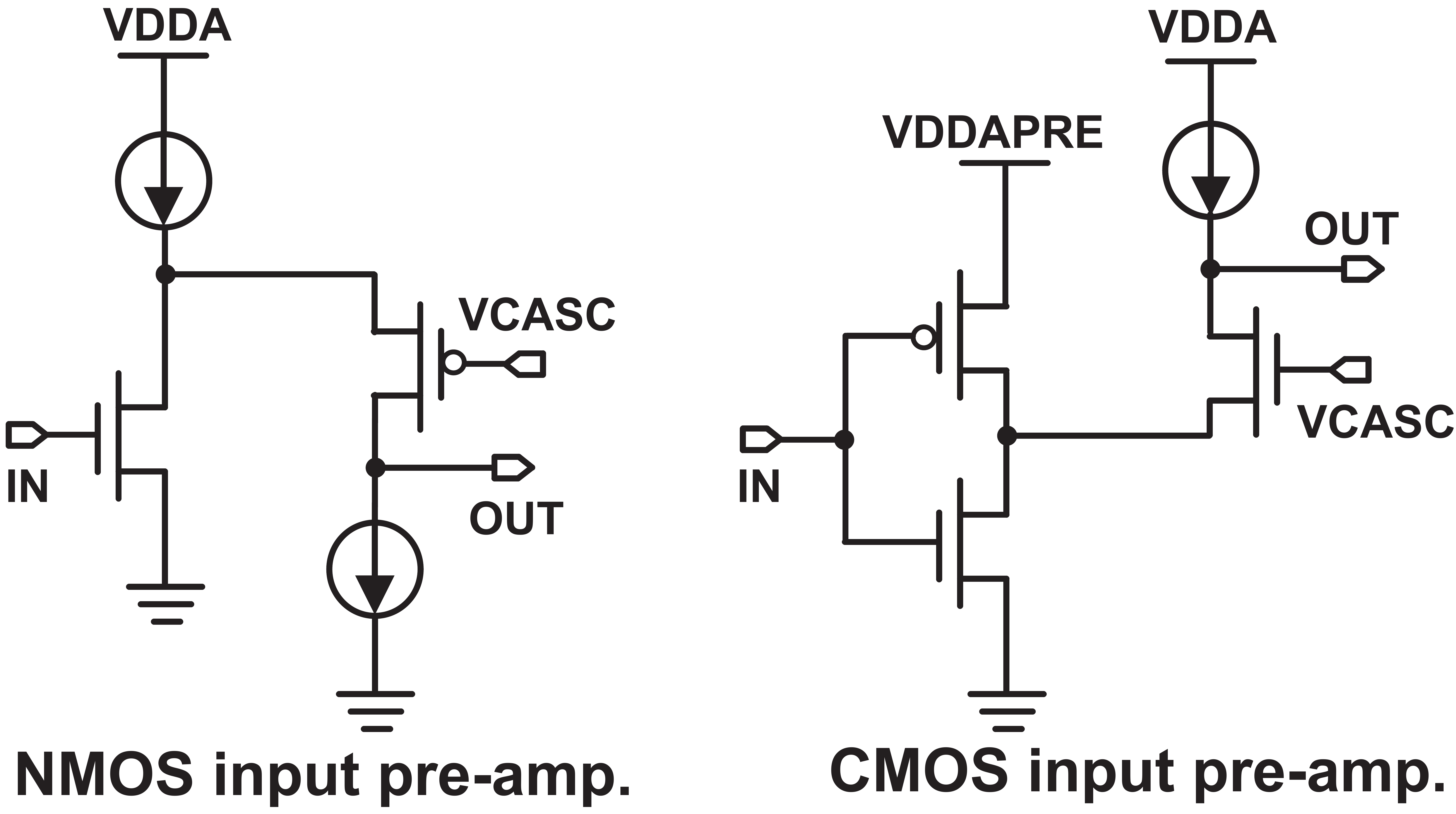}
      \label{fig:Amp}
    }
    \subfigure[]{
      \includegraphics[width=0.52\textwidth]{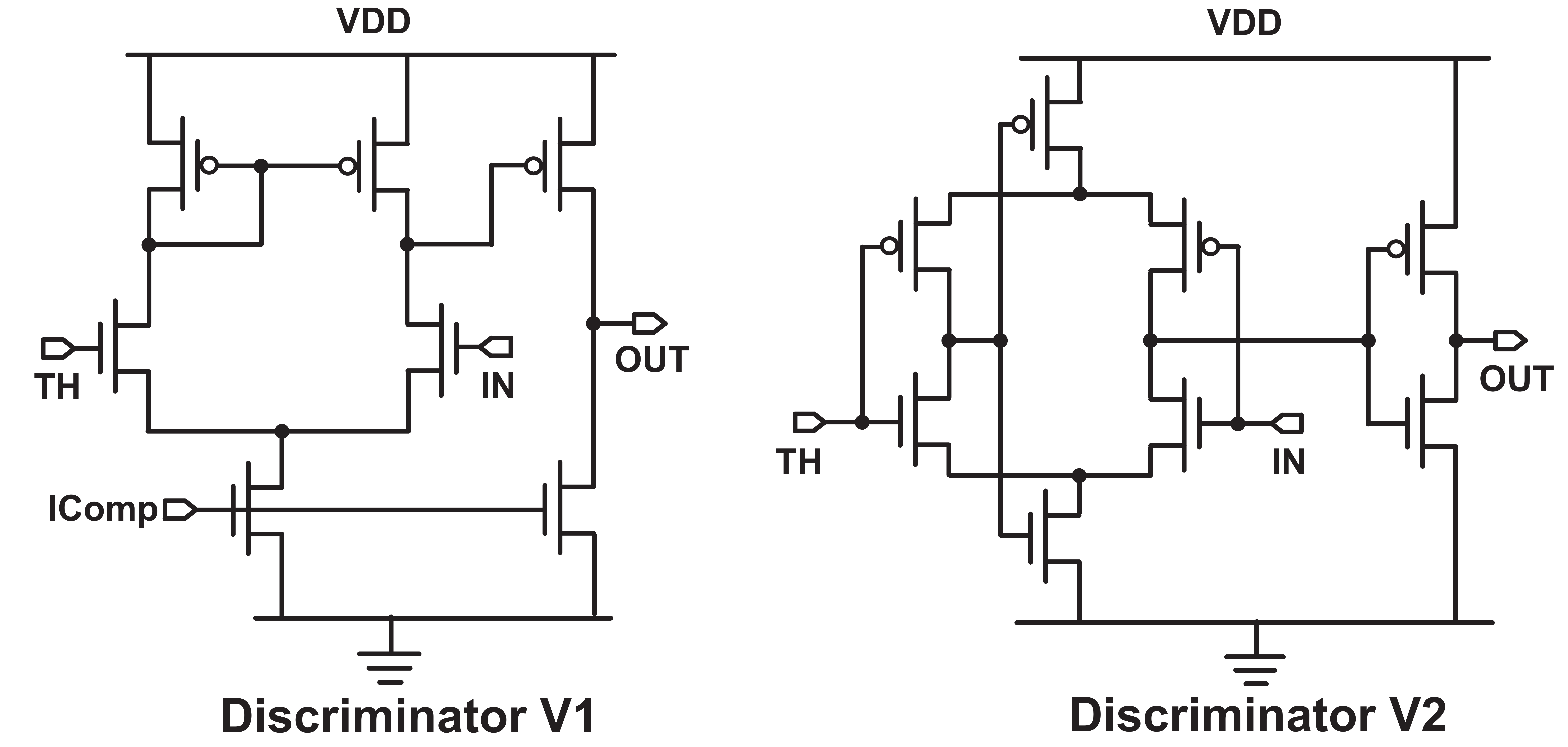}
      \label{fig:Dis}
    }
    \caption[]{Schematics of \subref{fig:Amp} pre-amplifiers and \subref{fig:Dis} discriminators implemented in LF-Monopix01.} 
    \label{fig:FE}
  \end{center}
\end{figure}

\paragraph{Pre-amplifier} There are two pre-amplifier designs in LF-Monopix01, and their schematics are given in Figure~\ref{fig:Amp}. Both pre-amplifiers use the single-ended folded cascode topology, and their transistor parameters were optimized to cope with the expected large detector capacitance ($\sim$~400~fF) in LF-Monopix01. The major difference between the two amplifiers is that one uses an NMOS transistor as the input device, whereas the other uses two complementary input transistors to increase the transconductance for a given bias current, and thus is expected to have better speed and noise performances~\cite{PreAmp_Yavuz_2016}. However, the bias of the CMOS input pre-amplifier relies on a separate analog power (VDDAPRE in Figure~\ref{fig:Amp}) provided by a regulator at the periphery, and the sensitivity of the circuit performance to this power line needs to be verified in a large scale sensor. 

\paragraph{Discriminator} There are also two discriminator designs in LF-Monopix01 (see Figure~\ref{fig:Dis}). The discriminator V1 has a typical two-stage open loop structure and it has been implemented in previous LFoundry prototypes. The discriminator V2 is a self-biased differential amplifier, followed by a CMOS inverter as the output stage. The latter exploits the complementary character of CMOS input devices for high gain and fast operation~\cite{CSDA_Bazes_1991}. 

\paragraph{Cross-talk minimization} As mentioned in Section~\ref{sec:ChallengeMonopix}, significant cross-talk may be introduced from the electronic layer to the sensing node via C$_{\text{pw}}$, mimicking signals. Therefore, the digital read-out logic employs a full custom design to minimize the area, and accordingly the capacitance of C$_{\text{pw}}$. In addition, the switching noise of some critical digital blocks were minimized. For instance, a current steering logic~\cite{CSL_Hiok_1997} is used for token propagation. Moreover, the readout of the memory cells adopts a source follower as the output stage, so that high current injection into the substrate of in-pixel electronic is avoided during the readout. In this prototype, another read-out concept was implemented by moving the read-out logic to the column end, forming a matrix of logic units. In this case, the drawbacks of the complex in-pixel circuitry are eliminated. However, the insensitive area is increased, and one-to-one connection from the sensor pixel to the corresponding logic unit is needed, which will complicate the layout design when scaling up the pixel array. 

\subsection{Chip architecture}
\label{sec:Chip}

\begin{figure}[!ht]
  \begin{center}
    \subfigure[]{
      \includegraphics[width=0.35\textwidth]{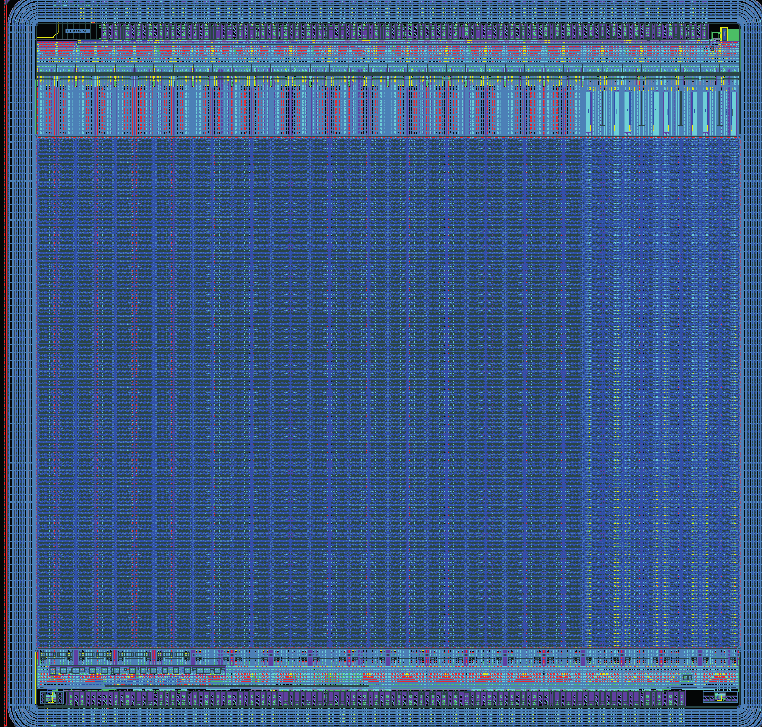}
      \label{fig:ChipLayout}
    }
    \subfigure[]{
      \includegraphics[width=0.5\textwidth]{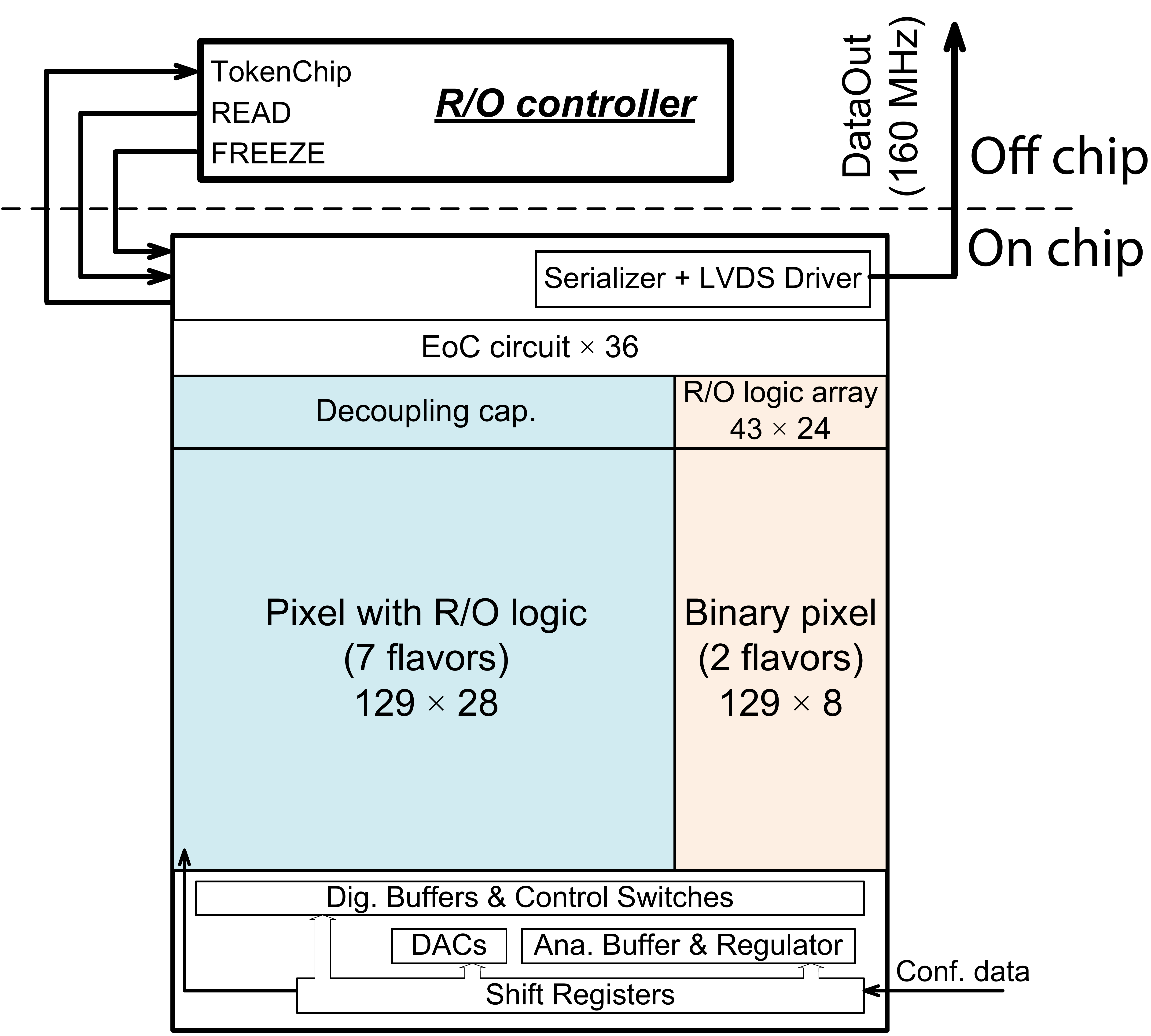}
      \label{fig:ChipBlock}
    }
    \caption[]{\subref{fig:ChipLayout} The layout and \subref{fig:ChipBlock} block diagram of LF-Monopix01.} 
    \label{fig:Chip}
  \end{center}
\end{figure}

The size of the LF-Monopix01 prototype is $\sim$~10~$\times$~10~mm$^{\text{2}}$. The chip layout, as well as the block diagram, are shown in Figure~\ref{fig:Chip}. The bottom part of the prototype includes the circuits that are used for chip bias, configuration and monitoring. The main body contains an array of 129~$\times$~36 pixels. In order to perform comparative studies, there are in total nine pixel flavors, each of them occupying four pixel columns. Seven flavors are pixels with in-pixel read-out logic. They differ from each other by different circuit block designs (i.e. pre-amplifier and discriminator) and layout implementations. The other two flavors are binary pixels with the read-out logic located at the column end. These pixels have the same pre-amplifier and discriminator design, but different output stages are used to drive the binary information to the column bus. On the top side, each pixel column interfaces with its own End-of-Column (EoC) circuitry. The EoC circuitry includes: 24 sense amplifiers to receive and temporally store the 24-bit data from the column bus, a gray counter running at 40~MHz to provide the timing information to the column, and the EoC logic which performs the column level read-out priority scan and data transmission. The data from the EoC circuitry is serialized on-chip and sent out by a LVDS driver with a bit rate of 160 Mbps. For design simplicity, the read-out controller will be implemented off-chip by the FPGA.

\section{Simulation results}
\label{sec:Sim}

The functionality and performance of LF-Monopix01 have been intensively simulated, and some selected results are presented in this section.

\paragraph{Gain and noise}

\begin{figure}[!ht]
  \begin{center}
    \subfigure[]{
      \includegraphics[width=0.37\textwidth]{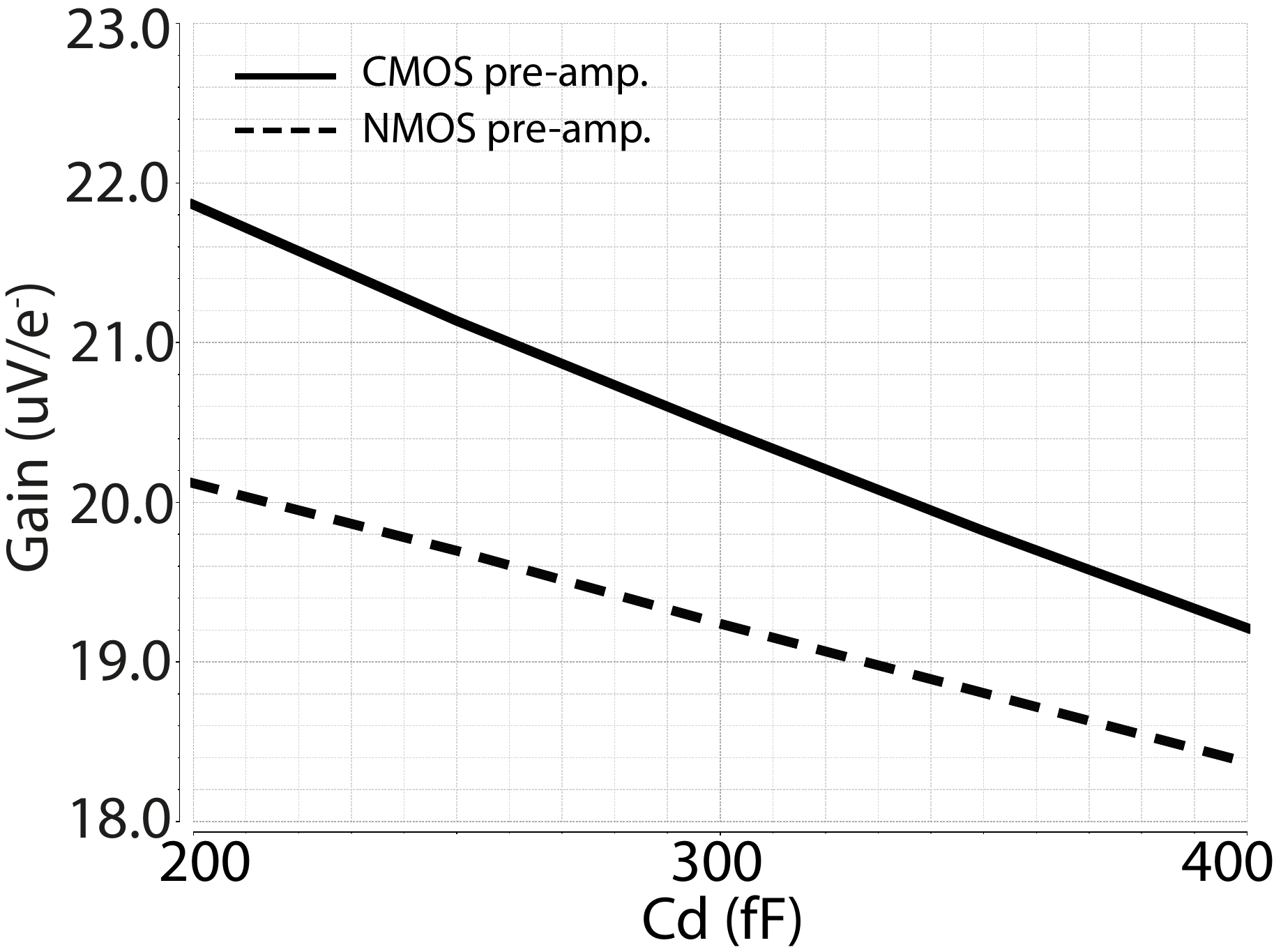}
      \label{fig:Gain}
    }
    \subfigure[]{
      \includegraphics[width=0.38\textwidth]{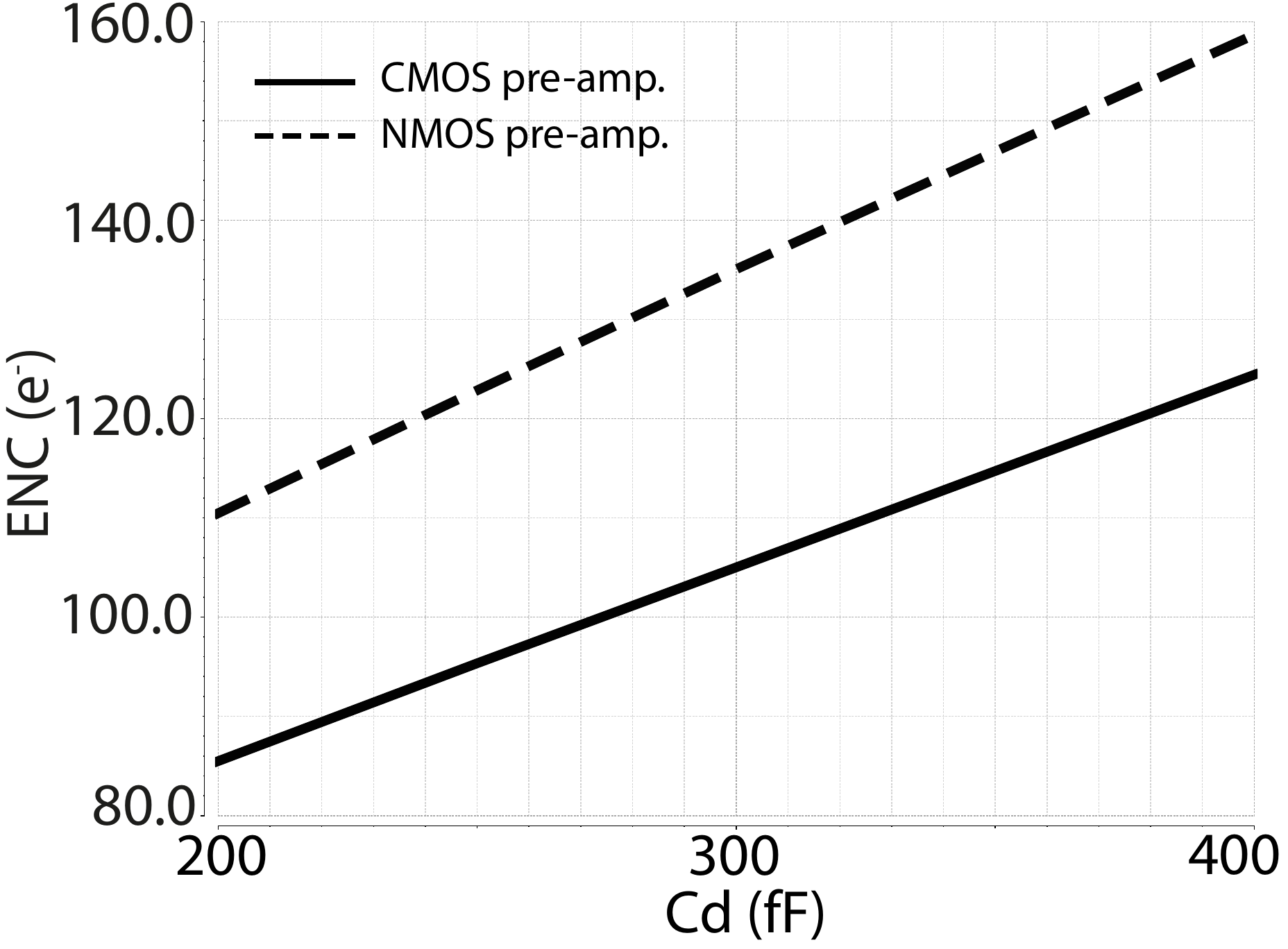}
      \label{fig:Noise}
    }
    \caption[]{Simulated \subref{fig:Gain} gain and \subref{fig:Noise} noise for the NMOS (dashed line) and CMOS (solid line) pre-amplifiers, as a function of detector capacitance.}
    \label{fig:GN} 
  \end{center}
\end{figure}

Figure~\ref{fig:GN} shows the simulated gain and noise of the pre-amplifier, as a function of detector capacitance. The input signal charge is 4000~e$^{-}$. The NMOS pre-amplifier (dashed line) has a bias current of $\sim$~17.5~$\mu$A. With 400~fF input capacitance, the gain is above 18~$\mu$V/e$^{-}$ and the equivalent noise charge (ENC) is below 160~e$^{-}$. The CMOS amplifier (solid line) has a slightly lower bias current of $\sim$~15~$\mu$A. Thanks to its complementary input transistors, the CMOS pre-amplifier has less noise and larger gain, as compared to the NMOS counterpart.

\paragraph{Time walk}

\begin{figure}[htbp]
\centering % \begin{center}/\end{center} takes some additional vertical space
\includegraphics[width=.48\textwidth]{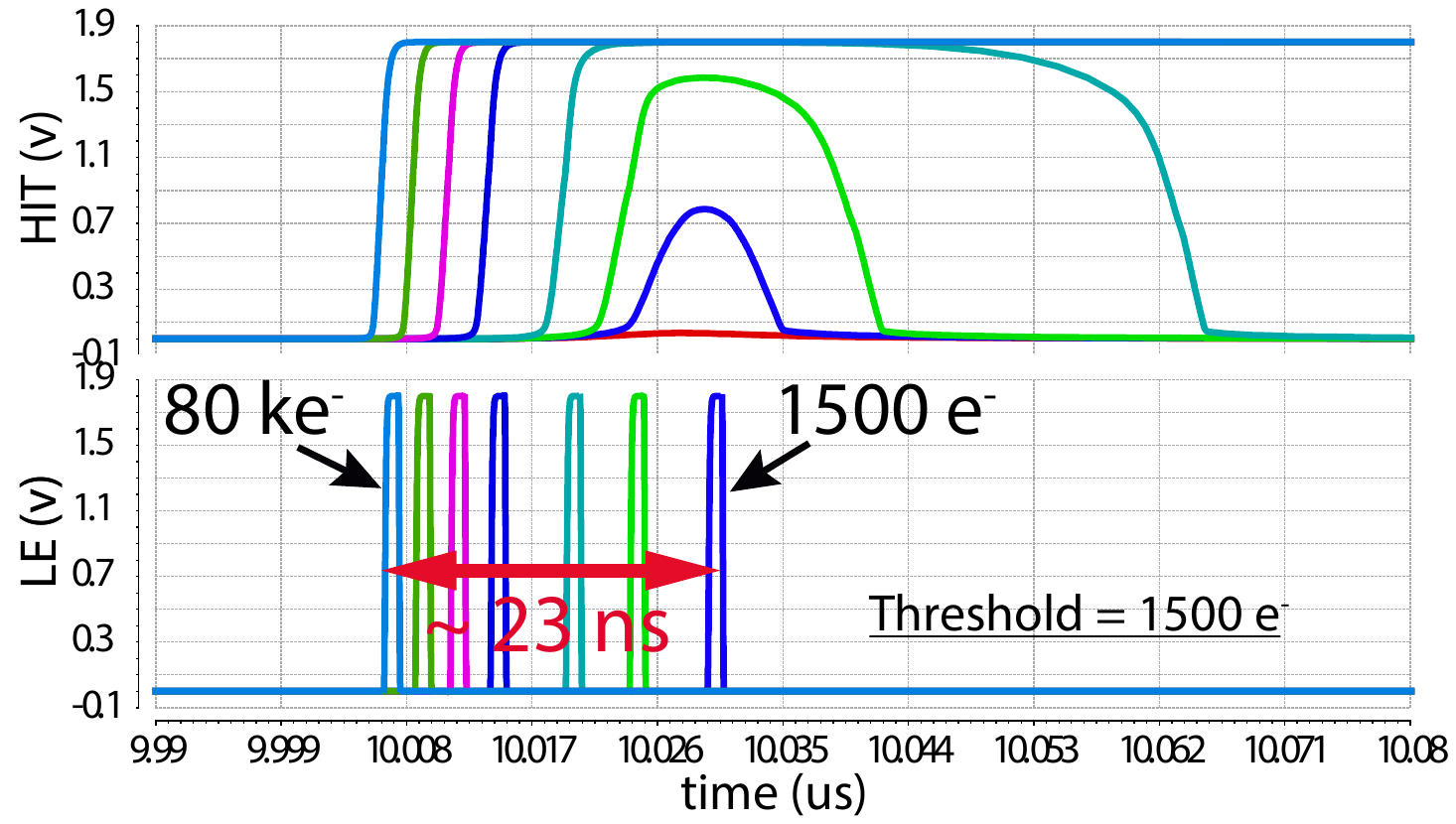}
\caption{\label{fig:TWNMOSV2} Time walk simulation for a pixel with NMOS pre-amplifier and discriminator V2.}
\end{figure}

Figure~\ref{fig:TWNMOSV2} shows the simulated waveforms of the discriminator output (``HIT'') and the LE pulse, with the input charge ranging from 1500~e$^{-}$ to 80~ke$^{-}$.  The simulated pixel uses the NMOS pre-amplifier and the discriminator V2. The detector capacitance is assumed to be 400~fF. The threshold of the discriminator is adjusted to be 1500~e$^{-}$, meaning that only signal charge beyond this value will fire the discriminator and generate the LE pulse. The resulting LE pulse indicates the arrival of the hit. The simulated time walk is $\sim$~23~ns ,which is the time difference between the two LE pulses that are generated from the ``small'' signal (1500~e$^{-}$) and a very large charge signal (80~ke$^{-}$) respectively.

\paragraph{Column simulation}

\begin{figure}[!ht]
  \begin{center}
    \subfigure[]{
      \includegraphics[width=0.3\textwidth]{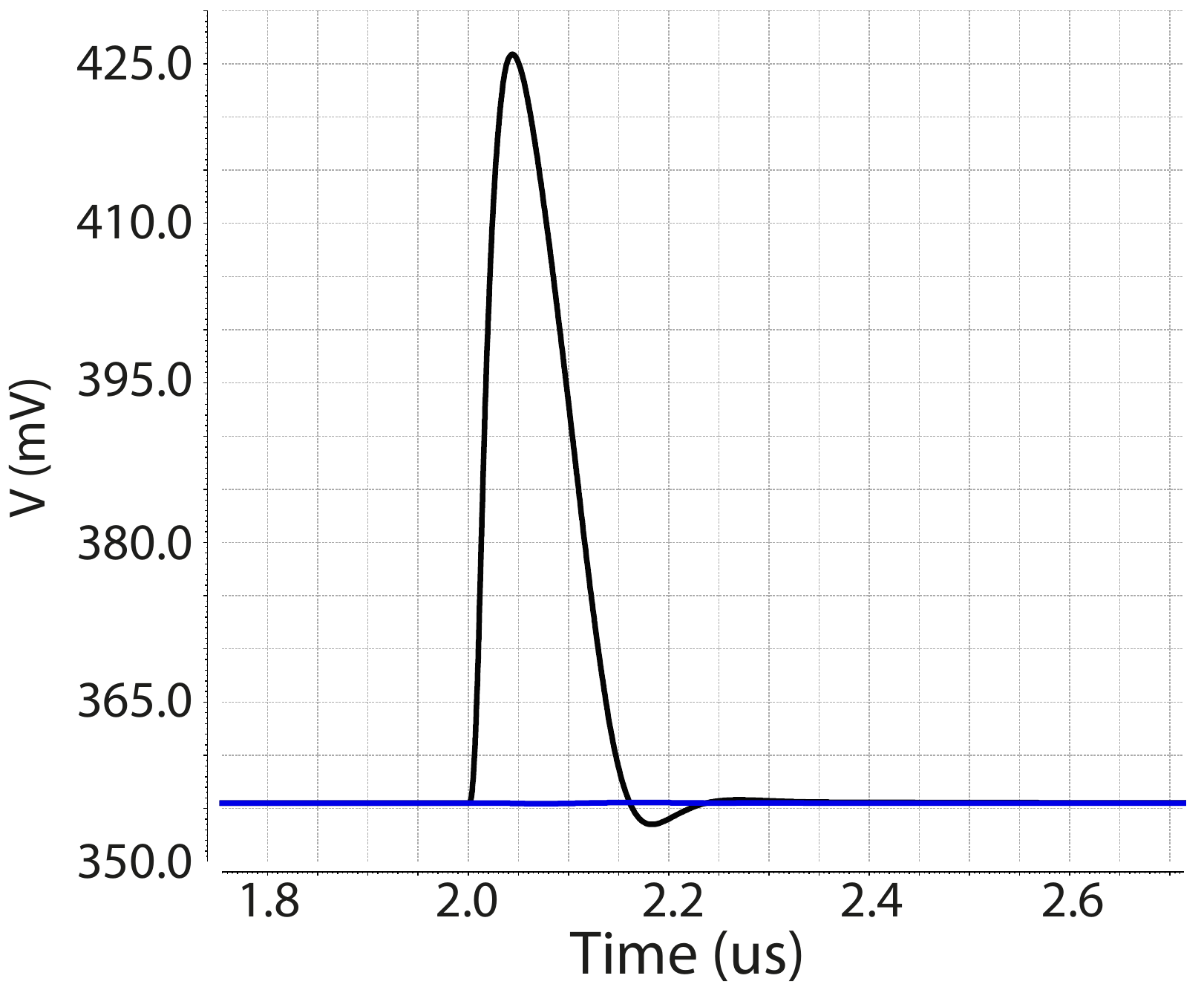}
      \label{fig:ColSim}
    }
    \subfigure[]{
      \includegraphics[width=0.3\textwidth]{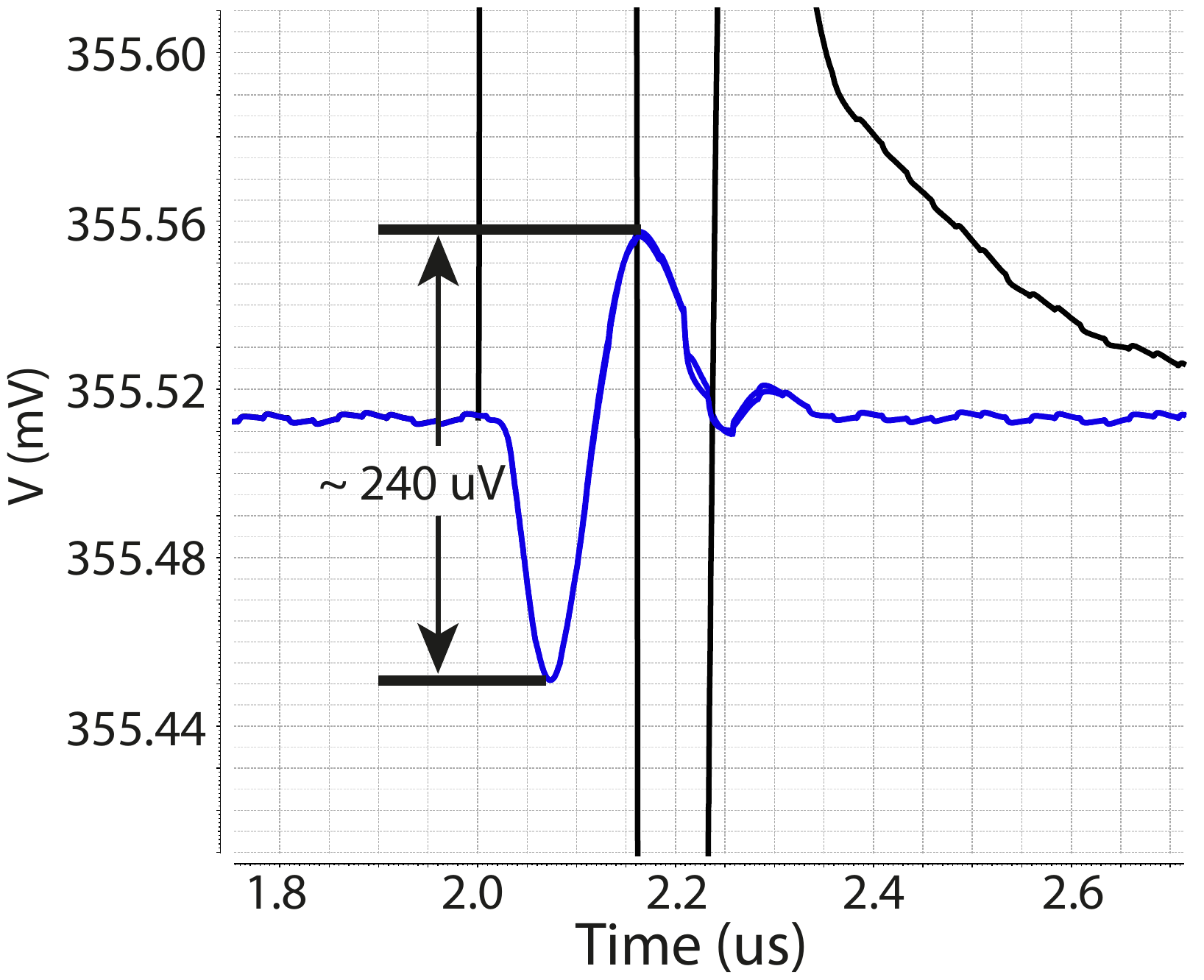}
      \label{fig:ColSimZoom}
    }
    \caption[]{\subref{fig:ColSim} Simulated pre-amplifier outputs of the hit pixel (black) and two neighboring pixels (blue) in the column. \subref{fig:ColSimZoom} The zoomed-in view around the baseline.} 
    \label{fig:ColPost}
  \end{center}
\end{figure}

The biasing and power/ground lines are distributed column-wise in the layout of the pixel array. Thus, post layout simulation of an entire pixel column was performed, in order to study the parasitic effects. The simulated column is composed of pixels with integrated read-out logic. Figure~\ref{fig:ColSim} shows the response of the pre-amplifier in the hit pixel, together with the responses of two neighboring pixels. As can be seen from the zoomed-in view around the baseline level (see Figure~\ref{fig:ColSimZoom}), the perturbation in the two neighboring pixels, caused by the hit pixel, is only $\sim$~240~$\mu$V in peak-to-peak value. This is equivalent to a charge injection less than 15~e$^{-}$. Moreover, the digital activities in the pixel have negligible influence on the analog part during the hit readout. 

\paragraph{Full chip simulation}

\begin{figure}[htbp]
\centering % \begin{center}/\end{center} takes some additional vertical space
\includegraphics[width=.45\textwidth]{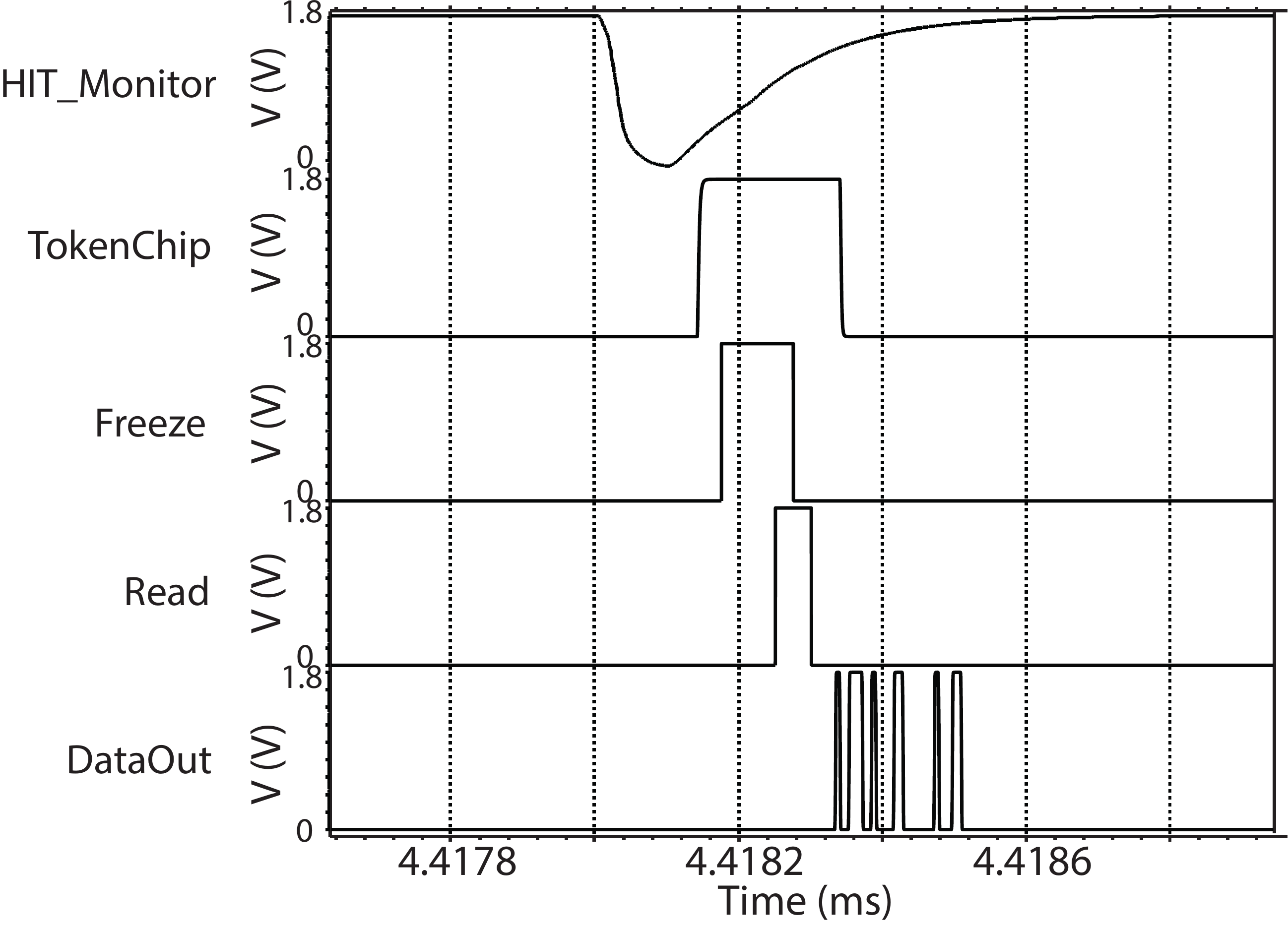}
\caption{\label{fig:ChipSim} Post layout chip simulation.}
\end{figure}

The full chip including the extracted parasitic was simulated with a FastSpice tool to verify the chip functionality. The read-out controller was realized by the Verilog-A model. The simulated waveforms, resulting from the signal injection into an arbitrary pixel, are shown in Figure~\ref{fig:ChipSim}. The firing of the pixel can be observed on a common monitoring line (\emph{HIT\_Monitor}). \emph{TokenChip}, \emph{Freeze} and \emph{Read} are the interfacing signals between the chip and the read-out controller as described in Section~\ref{sec:Pix}. \emph{DataOut} is the serialized data, including the hit timing and address.

\section{Conclusions}
\label{sec:Con}

Depleted CMOS pixels have shown their potential for the ATLAS ITk upgrade. One possible use case would be equipping the outer layers of the ITk upgrade with fast readout monolithic CMOS sensors. In this work, a depleted monolithic CMOS pixel sensor was designed in the LFoundry 150~nm CMOS technology. The prototype is capable of fast standalone readout by using a column drain architecture. Given by the measurement results from a previous prototype, the chip is foreseen to withstand the radiation levels expected at the outer layers of the ITk upgrade. The FE electronics were optimized in this prototype for a better timing performance, and a time walk value of $\sim$~23~ns was achieved in simulation. Negligible cross talk was seen when simulating one extracted pixel column with the SPICE-level accuracy. The functionality of the full chip, containing over two million transistors, was verified by post layout simulation using a FastSpice tool. The design was submitted in August 2016, and the delivery on silicon is expected in Q1 of 2017.

% \appendix
% \section{Some title}
% Please always give a title also for appendices.

% \acknowledgments


\end{document}